\documentclass[%
reprint,
amsmath,amssymb,
aps,
]{revtex4-2}

\usepackage{graphicx}
\usepackage{dcolumn}
\usepackage{bm}

\begin{document}


\title{Holographic torus correlators of stress tensor in $AdS_3/CFT_2$}

\author{Song He$^{1,2}$}
 \email{hesong@jlu.edu.cn}
\author{Yi Li$^{1}$}%
 \email{liyi@fudan.edu.cn}
\author{Yun-Ze Li$^{1}$}
 \email{lyz21@mails.jlu.edu.cn}
\author{Yunda Zhang$^{1}$}%
 \email{ydzhang21@mails.jlu.edu.cn}
\affiliation{%
 $^{1}$Center for Theoretical Physics and College of Physics, Jilin University, Changchun 130012, People's Republic of China
}%

\affiliation{%
 $^{2}$Max Planck Institute for Gravitational Physics (Albert Einstein Institute), Am M\"uhlenberg 1, 14476 Golm, Germany
}%

\date{\today}

\begin{abstract}
In the context of $\rm AdS_3/CFT_2$, we investigate holographic correlators of the stress tensor of a conformal field theory (CFT) on a torus in this work. To calculate the correlators of the stress tensor, we employ the Einstein-Hilbert theory of gravity and perturbatively solve Einstein's equation in the bulk. We offer an explicit prescription to develop a recurrence relation that makes it simple to compute higher point correlators. The correlators and the recurrence relation are found to be consistent with what is known in CFTs. Following the spirit of the proposed cutoff $\rm AdS$/$T\bar{T}$ CFT holography, we then expand our computation program to investigate holographic torus correlators at a finite cutoff in the $\rm AdS_3$. A parallel recurrence relation associated with higher point correlators can be obtained.
\end{abstract}

\maketitle


\section{\label{sec:introduction}Introduction}

Understanding nonperturbative effects in the presence of strong coupling is one of the most difficult problems in modern physics. Analytical results in strong coupling are rare and extremely difficult to obtain. A remarkable tool for strongly coupled QFT is offered by the holographic principle \cite{tHooft:1993dmi, Susskind:1994vu}. The Anti-de Sitter gravity/conformal field theory (AdS/CFT) correspondence \cite{Maldacena:1997re, Gubser:1998bc, Witten:1998qj} provides a rare window to gain analytical insight into strongly coupled physics. In the most helpful limit to exploit this correspondence, we use the weakly coupled bulk description to study the physics of strong coupling by performing gravitational perturbative calculations.

On the other hand, we are still far from harnessing the full computational power of AdS/CFT correspondence, which is particularly evident when considering the most fundamental observables of the CFT, namely the correlators of local operators. In particular, previous studies on holographic correlators of the stress tensor focus on CFTs with trivial topology with holographic computation done in the pure AdS space \cite{Liu:1998ty, DHoker:1999bve, Arutyunov:1999nw, Raju:2012zs}, even in $\text{AdS}_3/\text{CFT}_2$ \cite{Bagchi:2015wna}. Once we apply the variational principle to investigate the correlators in the CFTs with nontrivial topology, it amounts to solving Einstein's equation in a nontrivial background bulk geometry. While near-boundary solutions are well-understood \cite{Fefferman:1985ci, Henningson:1998gx, Skenderis:1999nb,deHaro:2000vlm,Fefferman:2007rka}, without the full symmetry of pure AdS the global boundary value problem is in general very difficult (see \cite{Graham:1991jqw,anderson2004structure, Anderson:2004yi,anderson2008einstein} for some discussion), even for linearized equations. Further, the higher-point stress tensor correlators in strongly coupled CFTs with nontrivial topology are still less known in the community. Explicit results from holographic computation are highly desirable.

In this letter, we tackle and solve this long-standing problem in $\text{AdS}_3/\text{CFT}_2$. We begin by computing the holographic torus correlators of stress tensor by solving Einstein's equation in the thermal AdS background. Then we propose a prescription, which in principle works for any Riemann surface, to compute the $n$-point correlators by deriving a recurrence relation. These results are consistent with the corresponding data in CFTs \cite{He:2020udl}. 

We further extend this program of computing holographic stress tensor correlators to the context of cutoff AdS/$T\Bar{T}$ CFT holography. It was proposed that for $T\bar{T}$ deformation \cite{Zamolodchikov:2004ce, Smirnov:2016lqw} of holographic CFTs, the holographic dual is to move the conformal boundary to a finite cutoff in the bulk AdS space \cite{McGough:2016lol}. Stress tensor correlators of $T\Bar{T}$-deformed CFT have been considered in the complex plane \cite{Kraus:2018xrn, Aharony:2018vux} and on the torus \cite{He:2020udl, He:2020cxp}, and holographic correlators in the complex plane were studied in \cite{Kraus:2018xrn, Li:2020pwa} following the spirit of cutoff AdS holography. Here we compute the holographic correlators on a torus at a finite cutoff in the thermal $\rm AdS_3$ and derive a recurrence relation similar to the case of CFTs.



\section{\label{sec:general formalism}Holographic setup}

For holographic computation, it's customary to work in the Fefferman-Graham coordinates near the conformal boundary, in which the bulk metric takes the form
\begin{align} \label{bulk metric FG series}
    ds^2 = \frac{dr^2}{r^2} + \frac{1}{r^2}g_{ij}(x,r) dx^i dx^j.
\end{align}
In dimension three, the Fefferman-Graham series of the metric truncates as \cite{Skenderis:1999nb}
\begin{equation} \label{FG series in 3D}
    g_{ij}(x,r)=g^{(0)}_{ij}(x) + g^{(2)}_{ij}(x) r^2 + g^{(4)}_{ij}(x) r^4
\end{equation}
and Einstein's equation is reduced to one equation that determines $g^{(4)}$ in terms of $g^{(0)}$ and $g^{(2)}$:
\begin{equation} \label{g4 by g0 and g2}
    g^{(4)}_{ij}=\frac{1}{4}g^{(2)}_{ik}g^{(0)kl}g^{(2)}_{lj}.
\end{equation}
and another two equations
\begin{align}
    {\nabla^{(0)}}^i g^{(2)}_{ij} &= \nabla^{(0)}_j {g^{(2)}}_i^i, \label{g2 conservation}\\
    {g^{(2)}}_i^i &= -\frac{1}{2}R[g^{(0)}] \label{g2 trace}
\end{align}
where the covariant derivative and raising (lowering) indices are all with respect to the metric $g^{(0)}$. If the holographic field theory lives on a cutoff surface $r=r_c$ as the boundary, the background metric of the field theory $\gamma$ is identified with the induced metric $h$ on the boundary by
\begin{align} \label{background metric and boundary metric}
    \gamma_{ij} = r_c^2 h_{ij}
\end{align}
and by taking the functional derivative of the gravity on-shell action with respect to the boundary metric, the one-point correlator of the stress tensor is identified with the Brown-York tensor on the boundary with proper counter-terms added \cite{Balasubramanian:1999re, Emparan:1999pm,deHaro:2000vlm}
\begin{align} \label{one-point correlator and BY tensor}
    \langle T_{ij} \rangle = -\frac{1}{8\pi G}(K_{ij} - K h_{ij} + h_{ij}).
\end{align}
Therefore, the pair of background metric and one-point correlator is identified with a near-boundary bulk geometry, expressed in terms of $g^{(0)}$ and $g^{(2)}$. From (\ref{g2 conservation}) and (\ref{g2 trace}) (or from the Gauss-Codazzi equation in general dimensions), we have the conservation equation
\begin{align} \label{one-point correlator conservation}
    \nabla^i \langle T_{ij} \rangle = 0,
\end{align}
and the trace relation
\begin{align} \label{one-point correlator trace relation}
    \langle T \rangle = \frac{R}{16\pi G} + 4\pi G r_c^2 (\langle T^{ij} \rangle \langle T_{ij} \rangle - \langle T \rangle^2).
\end{align}
For CFTs, the cutoff surface is pushed to the conformal boundary $r_c=0$, and the trace relation reduces to the Weyl anomaly
\begin{align} \label{Weyl anomaly}
    \langle T \rangle = \frac{R}{16\pi G}.
\end{align}
For $T\bar{T}$-deformed CFTs, the cutoff location is related to the $T\bar{T}$ deformation parameter by \cite{McGough:2016lol}
\begin{align} \label{cutoff location and deformation parameter}
    \mu = 16 \pi G r_c^2
\end{align}

To obtain multi-point correlators, it suffices to compute the varied one-point correlator by a variation of the boundary metric.
Equations (\ref{one-point correlator conservation}) and (\ref{Weyl anomaly})(or (\ref{one-point correlator trace relation})) serve as our basis for this computation, but they don't fully determine the one-point correlator, or in the bulk language, the near-boundary solution to Einstein's equation is (understandably) not unique. The remaining information, which in our case is two constants of integration, must be drawn from the global geometry of the bulk space by requiring that the near-boundary solution reconstructed from the one-point correlator can be extended to be a global solution. We will show in our case that this condition, dubbed as the global regularity condition, fixes the two constants.

There is another way to fix the two constants, which naturally leads to deriving a recurrence relation of the holographic correlators. The two constants can be viewed as one-point-averaged correlators, which in turn correspond to variations of lower point correlators with respect to global metric variations. On a Riemann surface, global metric variations are related to the differentiation with respect to the moduli. This idea has already been used to study Ward identities of stress tensor insertions \cite{Friedan:1986ua, Eguchi:1986sb}. In particular, let the global metric variation be $\delta\gamma_{\bar z\bar z}(z)=\alpha, \, \delta\gamma_{zz}(z)=\bar\alpha$ for a torus. To the first order in $\alpha$, the new metric is
\begin{align}
	ds^2=&dzd\bar z+\bar\alpha dz^2+\alpha d\bar z^2\notag,\\
	=&(1+\alpha+\bar\alpha)d(z+\alpha(\bar z-z))d(\bar z+\bar\alpha(z-\bar z)).
\end{align}
With a Weyl transformation of factor $1-\alpha-\bar{\alpha}$ and a change of coordinates
\begin{equation}
	z'=z+\alpha(\bar z-z),\ \ \bar z'=\bar z+\bar\alpha(z-\bar z)
\end{equation}
we get a torus with the Euclidean metric and a varied modular parameter
\begin{align}
    \tau' = \tau + \alpha (\bar{\tau} - \tau).
\end{align}
Therefore for the correlator of any operators collectively denoted by $O$ we have
\begin{align}
    (\bar{\tau}-\tau) \partial_\tau \langle O \rangle &= \mathcal{L}_{(z-\bar{z})\partial_z} \langle O \rangle
    + \int_{\text{T}^2} d^2z \big( \frac{\delta \langle O \rangle}{\delta \gamma_{\bar{z}\bar{z}}(z)} -  \frac{\delta \langle O \rangle}{\delta \gamma_{z\bar{z}}(z)} \big) \label{global metric variation and modular differentiation eqn 1},\\
    (\tau-\bar{\tau}) \partial_{\bar{\tau}} \langle O \rangle &= \mathcal{L}_{(\bar{z}-z)\partial_{\bar{z}}} \langle O \rangle
    + \int_{\text{T}^2} d^2z \big( \frac{\delta \langle O \rangle}{\delta \gamma_{zz}(z)} -  \frac{\delta \langle O \rangle}{\delta \gamma_{z\bar{z}}(z)} \big), \label{global metric variation and modular differentiation eqn 2}
\end{align}
where $\cal L$ denotes the Lie derivative. In the community, it is the first explicit realization of \cite{Friedan:1986ua, Eguchi:1986sb} which is the key point to obtain the holographic recurrence relation of higher point stress tensor correlators in the remaining part of this letter.

\section{\label{sec:conformal boundary}Torus correlators in holographic CFT}
We start by computing the holographic torus correlators of stress tensor on the conformal boundary (for CFTs) from the thermal $\rm AdS_3$. Other classical gravity saddles (real smooth ones) with the torus conformal boundary are classified in \cite{Maloney:2007ud} (first considered in \cite{Maldacena:1998bw}). They can all be obtained from the thermal $\rm AdS_3$ by modular transformations. The thermal $\rm AdS_3$ is a solid torus with the metric
\begin{equation}
	ds^2=d\rho^2 + \cosh^2\rho dt^2 + \sinh^2\rho d\phi^2, \label{thermal AdS3 metric 1}
\end{equation}
or in the form of the Fefferman-Graham series
\begin{equation}
	ds^2=\frac{dr^2}{r^2}+\frac{1}{r^2}\Big[dzd\bar z - r^2 \pi^2(dz^2+d\bar z^2) + r^4 \pi^4dzd\bar z\Big], \label{thermal AdS3 metric 2}
\end{equation}
with
\begin{align}
    r =\frac{1}{\pi e^\rho},\; z = \frac{\phi+i t}{2\pi},\; \bar{z} =\frac{\phi-it}{2\pi},
\end{align}
where $z,\bar{z}$ are doubly periodically identified $(z,\bar z)\sim (z+1,\bar z+1)\sim (z+\tau,\bar z+\bar\tau)$. The conformal boundary at $\rho=\infty$ or $r=0$, is a torus with two periods 1 and $\tau$, with the Euclidean metric $\gamma_{ij}dx^i dx^j=dz d\bar{z}$. We read off one-point correlators from the bulk geometry
\begin{equation} \label{1pt correlator thermal AdS3}
	\langle{T_{zz}}\rangle=-\frac{\pi }{8G},\ \ \langle{T_{\bar z\bar z}}\rangle=-\frac{\pi }{8G},\ \ \langle{T_{z\bar z}}\rangle=0.
\end{equation}
To compute the holographic correlators, we take a variation of the metric
\begin{align}
    \delta\gamma_{ij} dx^i dx^j = \epsilon f_{ij} dx^i dx^j.
\end{align}
and expand the variation of the one-point correlator in powers of the infinitesimal parameter $\epsilon$
\begin{align}
    \sum_{n=1}^\infty \epsilon^n T^{[n]}_{ij}.
\end{align}
From (\ref{one-point correlator conservation}) and (\ref{Weyl anomaly}), we can order-by-order solve $T^{[n]}_{ij}$ to compute $n+1$ point correlators. For the first order, we find
\begin{align} \label{T1 conformal boundary}
    T^{[1]}_{z\bar{z}} =& \frac{1}{16\pi G}(-2\pi^2 (f_{zz} + f_{\bar{z}\bar{z}}) + \partial_{\bar{z}}^2 f_{zz} - 2\partial_z\partial_{\bar{z}} f_{z\bar{z}} + \partial_z^2 f_{\bar{z}\bar{z}}), \nonumber\\
    T^{[1]}_{zz}(z) &= \frac{1}{16\pi G}
    \Big[ (-\partial_z\partial_{\bar{z}}f_{zz} + 2 \partial_z^2 f_{z\bar{z}})(z) \nonumber\\
    &+ \frac{1}{\pi}\int_{\text{T}^2} d^2w G_\tau(z-w) (-4\pi^2 \partial_w - \partial_w^3) f_{\bar{z}\bar{z}}(w) + C^{[1]}
    \Big],\notag\\
    T^{[1]}_{\bar{z}\bar{z}}(z) &= \text{c.c. of} \; T^{[1]}_{zz}(z)
\end{align}
where ``c.c." denotes complex conjugate, $C^{[1]},\bar{C}^{[1]}$ are constants of integration, and
\begin{align}
    G_\tau(z) = \zeta_\tau (z) - 2\zeta_\tau\big(\frac{1}{2}\big) z + \frac{2\pi i}{\text{Im}\tau}\text{Im}z
\end{align}
is a Green's function on a torus (see the appendix \ref{A} for details) with $\zeta_\tau (z)$ being the Weierstrass Zeta function. 

As discussed in the previous section, any choice of the constants corresponds to a near-boundary solution to Einstein's equation in its Fefferman-Graham coordinates. The Fefferman-Graham coordinates of the varied bulk metric may differ from the $\rho,\phi,t$ or $r,z,\bar{z}$ coordinates of the solid torus, but we can make them coincide in the region $\rho \in (0,\infty)$ by a boundary preserving diffeomorphism. In other words, a generic bulk metric solution in the region $\rho \in (0,\infty)$ for a varied boundary metric is given by a Fefferman-Graham series in $\rho,\phi,t$ from (\ref{T1 conformal boundary}), plus a change by a boundary-preserving diffeomorphism. The global regularity condition in the present case is the metric must be regular at $\rho=0$. Leaving details of computation to the appendix \ref{B}, to the first order we have
\begin{align} \label{global regularity condition thermal AdS3}
    &\int_{\text{T}^2} d^2z [g^{(2)[1]}_{zz} - g^{(2)[1]}_{\bar{z}\bar{z}} + 2\pi^2 (g^{(0)[1]}_{zz} - g^{(0)[1]}_{\bar{z}\bar{z}})] = 0, \notag\\
    &\int_{\text{T}^2} d^2z [g^{(2)[1]}_{zz} + 2g^{(2)[1]}_{z\bar{z}} + g^{(2)[1]}_{\bar{z}\bar{z}}] = 0
\end{align}
where $g^{(0)[1]}$ and $g^{(2)[1]}$ are the first order variations of $g^{(0)}$ and $g^{(2)}$ respectively. This condition determines the constants as
\begin{align}
    C^{[1]} = \frac{4\pi^2}{\text{Im}\tau} \int_{\text{T}^2} d^2z f_{\bar{z}\bar{z}}, \;
    \bar{C}^{[1]} = \frac{4\pi^2}{\text{Im}\tau} \int_{\text{T}^2} d^2z f_{zz}
\end{align}
and all two-point correlators are obtained
\begin{align} \label{2pt correlator thermal AdS3}
    \langle T_{zz}(z) T_{zz}(w)\rangle &= \frac{1}{32\pi^2 G}\Big(\wp_\tau^{''}(z-w) + 4\pi^2 \wp_\tau(z-w) + 8\pi^2 \zeta_\tau\big(\frac{1}{2}\big)\Big)
\end{align}
where $\wp_\tau(z) = -\zeta_\tau^{'}(z)$ is the Weierstrass $P$ function. For simplicity, we have only shown the correlator of $T_{zz}$, since other components can be determined from it by the Ward identity of conservation.

Alternatively, the constants can be derived from (\ref{global metric variation and modular differentiation eqn 1}) and (\ref{global metric variation and modular differentiation eqn 2}) in the form of a recurrence relation. We begin by turning on a variation of $\gamma_{\bar{z}\bar{z}}=F$ of the Euclidean metric while keeping other components fixed, then we get the holographic Virasoro Ward identity \cite{Polyakov:1987zb,Banados:2004nr} from (\ref{one-point correlator conservation}) and (\ref{Weyl anomaly})
\begin{align} \label{holographic virasoro Ward identity}
    \partial_{\bar{z}} \langle T_{zz} \rangle - 2 \partial_z F \langle T_{zz} \rangle - F \partial_z \langle T_{zz} \rangle + \frac{1}{16\pi G} \partial_z^3 F = 0.
\end{align}
Taking the $n-$th functional derivative with respect to $F$ and evaluating at $F=0$, we find
\begin{align}
    &\partial_{\bar{z}} \langle T_{zz}(z) T_{zz}(z_1) \ldots T_{zz}(z_n) \rangle \nonumber\\
    &- \sum_{i=1}^n \partial_{z}\delta(z-z_i) \langle T_{zz}(z) T_{zz}(z_1) \ldots T_{zz}(z_{i-1})T_{zz}(z_{i+1}) \ldots T_{zz}(z_n) \rangle \nonumber\\
    &- \frac{1}{2} \sum_{i=1}^n \delta(z-z_i) \partial_z \langle T_{zz}(z) T_{zz}(z_1) \ldots T_{zz}(z_{i-1})T_{zz}(z_{i+1}) \ldots T_{zz}(z_n) \rangle \nonumber\\
    &+ \frac{1}{32\pi G} \delta_{n,1} \partial_z^3 \delta(z-z_1)= 0.
\end{align}
Solving with Green's function on the torus, we have
\begin{align}
    &\langle T_{zz}(z) T_{zz}(z_1) \ldots T_{zz}(z_n) \rangle \notag\\
    &=  \frac{1}{\pi} \sum_{i=1}^n \Big[\partial_z G_\tau(z-z_i) \langle T_{zz}(z_1) \ldots T_{zz}(z_n) \rangle \nonumber\\
    &- \frac{1}{2} G_\tau(z-z_i) \partial_{z_i} \langle T_{zz}(z_1) \ldots T_{zz}(z_n) \rangle \Big] \notag\\
    &- \frac{1}{32\pi^2 G}\delta_{n,1}\partial_z^3 G_\tau(z-z_1) \nonumber\\
    &+ \frac{1}{\text{Im}\tau} \int_{\text{T}^2} d^2v \langle T_{zz}(v) T_{zz}(z_1) \ldots T_{zz}(z_n) \rangle.
\end{align}
The last term of the one-point-averaged correlator, corresponding to the constants of integration, can be obtained from (\ref{global metric variation and modular differentiation eqn 1}) by setting $O=T_{zz}(z_1)\ldots T_{zz}(z_n)$ with the last term vanishing for CFTs. Then we obtain the recurrence relation 
\begin{align} \label{recurrence relation conformal boundary}
    &\langle T_{zz}(z) T_{zz}(z_1) \ldots T_{zz}(z_n) \rangle =\nonumber\\
    &-i \partial_\tau \langle T_{zz}(z_1) \ldots T_{zz}(z_n) \rangle + \frac{1}{32\pi^2 G} \delta_{n,1}\wp_\tau^{''}(z-z_1)\notag\\
    &- \frac{1}{2\pi} \sum_{i=1}^n \Big[ 2(\wp_\tau(z-z_i)+2\zeta_\tau\big(\frac{1}{2}\big)) \langle T_{zz}(z_1) \ldots T_{zz}(z_n) \rangle \notag\\
    &+ (\zeta_\tau(z-z_i) - 2\zeta_\tau\big(\frac{1}{2}\big)(z-z_i))\partial_{z_i}\langle T_{zz}(z_1) \ldots T_{zz}(z_n) \rangle \Big],
\end{align}
which is consistent with field theory derivation offered by \cite{He:2020udl}. The recurrence relation recovers the two-point correlators for thermal $\rm AdS_3$ (\ref{2pt correlator thermal AdS3}) and provides an efficient way to compute higher-point correlators. For example, the three-point correlator is
\begin{align}
    &\langle T_{zz}(z_1) T_{zz}(z_2) T_{zz}(z_3)\rangle = -\frac{1}{64\pi^3 G} \nonumber\\
    &\Big[12 \wp_\tau(z_1-z_2)\wp_\tau(z_2-z_3)\wp_\tau(z_3-z_1) \nonumber\\
    &+ 4\pi^2 (\wp_\tau(z_1-z_2)\wp_\tau(z_2-z_3) + \wp_\tau(z_2-z_3)\wp_\tau(z_3-z_1) \nonumber\\
    &+ \wp_\tau(z_3-z_1)\wp_\tau(z_1-z_2)) + (16\pi^2 \zeta_\tau(\frac{1}{2}) - g_{2,\tau}) \nonumber\\
    &(\wp_\tau(z_1-z_2)+\wp_\tau(z_2-z_3)+\wp_\tau(z_3-z_1))\Big] + C_{TTT,\tau},
\end{align}
which we have put in a symmetric form by analyzing the pole structure of the expression obtained from the recurrence relation (\ref{recurrence relation conformal boundary}). In the derivation, we used the identity
\begin{align}
 \wp_\tau^{''}(z) &= 6 \wp_\tau(z)^2 - \frac{g_{2,\tau}}{2} \nonumber\\
 g_{2,\tau} &= 60\sum_{(m,n)\neq (0,0)} \frac{1}{(m+n\tau)^4}
\end{align}
and $C_{TTT,\tau}$ is a constant (of no simple expression known to us) that can be obtained by evaluating the expression at three given points. The recurrence relation can be used to compute the correlators for any gravity saddle when it dominates in the path integral or to compute the exact correlators from a full partition function if it's available, though this question is much more subtle and difficult, for example, see \cite{Yin:2007gv, Maloney:2007ud}.

\section{\label{finite cutoff}$T\bar{T}-$ deformed Torus correlators}
In this section, we compute the holographic stress tensor correlators on a torus at a finite cutoff (for $T\bar{T}$-deformed CFTs). We start by embedding a torus as a cutoff surface into the thermal $\rm AdS_3$
\begin{align}
    ds^2 = d\rho^2 + \pi^2 e^{2\rho} [dZ d\bar{Z} - e^{-2\rho}(dZ^2 + d\bar{Z}^2) + e^{-4\rho}dZ d\bar{Z}].
\end{align}
By taking the periods in $Z$ to be 1 and $\Omega = \frac{\tau + e^{-2\rho_c}\bar{\tau}}{1+e^{-2\rho_c}}$ and defining the coordinates for torus
\begin{align}\label{z-Z trans}
    z = \frac{Z-e^{-2\rho_c}\bar{Z}}{1-e^{-2\rho_c}}, \;\bar{z} = \frac{\bar{Z}-e^{-2\rho_c}Z}{1-e^{-2\rho_c}}
\end{align}
and the Fefferman-Graham radial coordinate
\begin{align}
    r = \frac{1}{\pi e^\rho (1-e^{-2\rho_c})},
\end{align}
we get a torus at $\rho=\rho_c$ with periods $1$ and $\tau$ in the $z,\bar{z}$ coordinates, and a field theory background metric $\gamma_{ij} dx^i dx^j = r_c^2 h_{ij} dx^i dx^j = dz d\bar{z}$.
We read off one-point correlators from the bulk geometry
\begin{align} \label{1pt correlator cutoff thermal AdS3}
    \langle T_{zz} \rangle &= -\frac{\pi}{8 G} \frac{1-e^{-2\rho_c}}{1+e^{-2\rho_c}}, \;
    \langle T_{\bar{z}\bar{z}} \rangle = -\frac{\pi}{8 G} \frac{1-e^{-2\rho_c}}{1+e^{-2\rho_c}} \nonumber\\
    \langle T_{z\bar{z}} \rangle &= \frac{\pi}{8 G} \frac{e^{-2\rho_c}-e^{-4\rho_c}}{1+e^{-2\rho_c}}.
\end{align}
As in the previous section, we can solve the varied one-point correlator order by order from (\ref{one-point correlator conservation}) and (\ref{one-point correlator trace relation}) for a variation of the field theory metric $\epsilon f_{ij} dx^i dx^j$. Leaving the computation of $T^{[1]}_{ij}$ to the appendix \ref{C}, we obtain the two-point correlator
\begin{align} \label{2pt correlator cutoff thermal AdS3}
    &\langle T_{zz}(z) T_{zz}(w)\rangle \notag\\
    &= \frac{1}{16\pi G} \Big\{ \frac{1}{(1+e^{-2\rho_c})^4} \Big[2\pi(\wp_\Omega(Z-W) + e^{-8\rho_c}\overline{\wp_\Omega(Z-W)}) \nonumber\\
    &+\frac{1}{2\pi}(\wp_\Omega^{''}(Z-W) + e^{-8\rho_c}\overline{\wp_\Omega^{''}(Z-W)}) \nonumber\\
    &+ 4\pi(\zeta_\Omega(\frac{1}{2})+e^{-8\rho_c}\zeta_{\bar{\Omega}}(\frac{1}{2}))\Big] \nonumber\\
    &-\frac{1}{(1-e^{-4\rho_c})^3}[2\pi^2 e^{-2\rho_c}(1-e^{-2\rho_c})^2(1+e^{-4\rho_c}) \nonumber\\
    &+ (2e^{-2\rho_c}-3e^{-6\rho_c}+2e^{-10\rho_c})\partial_z^2 \notag\\
    &-(e^{-4\rho_c}+e^{-8\rho_c})\partial_z\partial_{\bar{z}} + e^{-6\rho_c}\partial_{\bar{z}}^2]\delta(z-w)
    \Big\}.
\end{align}

To relate to the $T\bar{T}$-deformed CFT, we have $\rho_c= \sinh^{-1}\frac{1}{2\pi r_c}$ and $r_c$ in turn is related to the $T\bar{T}$ deformation parameter $\mu$ by (\ref{cutoff location and deformation parameter}).
As a cross-check, we can compute the generating functional $I$ from the one-point correlators (\ref{1pt correlator cutoff thermal AdS3}) (as a special case of (\ref{global metric variation and modular differentiation eqn 1}) and (\ref{global metric variation and modular differentiation eqn 2})) by
\begin{align}
    i \partial_\tau I &= \langle T_{zz}\rangle - \langle T_{z\bar{z}}\rangle \nonumber\\
    -i\partial_{\bar{\tau}} I &= \langle T_{\bar{z}\bar{z}}\rangle - \langle T_{z\bar{z}}\rangle
\end{align}
We obtain
\begin{align}
    I &= \frac{i\pi}{8G}(1-e^{-2\rho_c})(\tau-\bar{\tau}) \nonumber\\
      &= \frac{i}{\mu} (\tau-\bar{\tau}) \Big( \sqrt{1+\frac{\pi \mu c}{6}} - 1 \Big)
\end{align}
It satisfies the $T\bar{T}$ flow equation for partition function \cite{Cardy:2018sdv, Datta:2018thy, He:2020cxp} with CFT limit $I=\frac{i\pi}{12}c(\tau-\bar{\tau})$, cf. equation (2.14) in \cite{Datta:2018thy} with the identification of the deformation parameter $\mu = -2\lambda^2$.

A recurrence algorithm to compute higher-point correlators can be derived in a way similar to the previous section, but it does not have a simple form like (\ref{recurrence relation conformal boundary}). So we leave it to the appendix \ref{C}.

\section{\label{conclusion}Conclusions and perspectives}

In this letter, we investigate the holographic torus correlators of the stress tensor on the conformal boundary (for CFTs) and at a finite cutoff (for $T\bar{T}$-deformed CFTs). First, a direct calculation is provided by solving Einstein's equation with a torus boundary, and then we obtain a holographic recurrence algorithm to calculate the higher point correlators of the stress tensor. The resulting recurrence relation for holographic CFTs is identical to that found in CFTs. The recurrence relation in holographic $T\bar{T}$-deformed CFTs can be also obtained. The recurrence algorithm has a natural generalization to higher genus Riemann surfaces.

It's interesting to extend our computation of holographic correlators to other operators and to higher dimensions. Exact results will contribute to the understanding of CFTs of non-trivial topology, in terms of OPEs, conformal blocks \cite{Hadasz:2009db, Kraus:2017ezw, Gobeil:2018fzy, Alday:2020eua, Karlsson:2022osn, Pavlov:2023asi} and possible bootstrap programs \cite{Iliesiu:2018fao}. With a proper recipe of analytic continuation to the Minkowski signature, we can also obtain exact results for holographic transport coefficients, as was done in \cite{Policastro:2001yc} and numerous following works.



\begin{acknowledgments}
We want to thank Bin Chen, Cheng Peng, Hao-Yu Sun, Jie-Qiang Wu, Jia Tian, and Xi-Nan Zhou for useful discussions related to this work. S.H. also would like to appreciate the financial support from Jilin University and the Max Planck Partner group, as well as the Natural Science Foundation of China Grants No.~12075101, No.~12235016.
\end{acknowledgments}

\appendix

\section{Green's function on torus}\label{A}
In this appendix, we briefly 
introduce the Green's functions on a torus with modular parameter $\tau$. The defining equations for the Green's functions $G_\tau(z,w)$ and $\tilde{G}_\tau(z,w)$ are
\begin{align}
    \frac{1}{\pi}\partial_{\bar{z}} G_\tau(z,w) = \delta(z,w) - \frac{1}{\text{Im}\tau}
\end{align}
and
\begin{align}
    \frac{1}{\pi}\partial_z\partial_{\bar{z}} \tilde{G}_\tau(z,w) = \delta(z,w) - \frac{1}{\text{Im}\tau},
\end{align}
where $\delta(z,w)$ is the delta function with respect to the measure $d^2 z = \frac{i}{2}dz\wedge d\bar{z}$. Green's functions can be rewritten as $G_\tau(z-w)$ and $\tilde{G}_\tau(z-w)$ by translational invariance.

For Green's function $G_\tau(z)$, since it takes the form $\frac{1}{z}$ in the complex plane, it is tempting to represent Green's function on the torus as a formal series
\begin{align}
 \sum_{(m,n)\in \mathbb{Z}^2} \frac{1}{z-(m+n\tau)}   
\end{align}
with manifest double periodicity. To make the formal series convergent, it's natural to add the holomorphic terms $\sum\limits_{(m,n)\in \mathbb{Z}^2 / (0,0)} \frac{1}{(m+n\tau)} - \frac{z}{(m+n\tau)^2}$ and we obtain the Weierstrass Zeta function
\begin{align}
    \zeta_\tau(z) = \frac{1}{z} + &\sum_{(m,n)\in \mathbb{Z}^2 / (0,0)} \Big(\frac{1}{z-(m+n\tau)} \nonumber\\
    &+ \frac{1}{(m+n\tau)} + \frac{z}{(m+n\tau)^2}\Big).
\end{align}
It's straightforward to prove $\zeta_\tau(z+1) - \zeta_\tau(z)$ and $\zeta_\tau(z+\tau) - \zeta_\tau(z)$ are independent on $z$, so we can restore the double periodicity by adding a linear function, obtaining the Green's function on the torus
\begin{align}
    G_\tau(z) = \zeta_\tau (z) - 2\zeta_\tau(\frac{1}{2}) z + \frac{2\pi i}{\text{Im}\tau}\text{Im}z.
\end{align}

Similarly, we have
\begin{align}
    \tilde{G}_\tau(z) = \log(|\sigma_\tau(z)|^2) - \zeta_\tau(\frac{1}{2}) z^2 - \overline{\zeta_\tau(\frac{1}{2})} \bar{z}^2 - \frac{2 \pi}{\text{Im}\tau} (\text{Im}z)^2
\end{align}
where
\begin{align}
    \sigma_\tau(z) = z \prod_{(m,n)\in \mathbb{Z}^2/(0,0)} \Big(1-\frac{z}{m+n\tau}\Big) e^{\frac{z}{m+n\tau} + \frac{z^2}{2(m+n\tau)^2}}
\end{align}
is the Weierstrass sigma function, with its log derivative being the Weierstrass zeta function.

\section{Global regularity condition}\label{B}
As discussed in section \ref{sec:conformal boundary}, we can make the Fefferman-Graham coordinates of the varied bulk metric (Fefferman-Graham coordinates always exist near the conformal boundary, see \cite{Fefferman:2007rka} for example) coincide with the torus coordinates $\rho,\phi,t$ by a diffeomorphism in the region $\rho \in (0,\infty)$. So the varied bulk metric in this region is given by a Fefferman-Graham series in $\rho,\phi,t$, determined from (\ref{T1 conformal boundary}), plus a change by a boundary preserving diffeomorphism. We characterize the diffeomorphism by a vector expanded in powers of $\epsilon$
\begin{align} \label{diffeomorphism vector}
    V = \sum_{n=1}^\infty \epsilon^n V^{[n]}
\end{align}
To the first order, the varied bulk metric is given by
\begin{align} \label{Perturbed metric first order in epsilon}
    ds^2 =& (1+\epsilon {\cal L}_{V^{[1]}})(d\rho^2 + \cosh^2 \rho dt^2 + \sinh^2\rho d\phi^2) \nonumber\\
    &+ \epsilon g^{FG[1]}_{ij} dx^i dx^j.
\end{align}
where $g^{FG[1]}$ is the first-order variation of the bulk metric in its Feffermen-Graham coordinates, given by
\begin{align} \label{delta gFG1}
    {g^{FG[1]}}_{zz} =& {g^{(2)[1]}}_{zz} - e^{-2\rho} {g^{(2)[1]}}_{z\bar{z}} + \pi^2 e^{2\rho} 
    g^{(0)[1]}_{zz} \nonumber\\
    &- \pi^2 e^{-2\rho} g^{(0)[1]}_{\bar{z}\bar{z}}, \nonumber\\
    {g^{FG[1]}}_{z\bar{z}} =& -\frac{1}{2} e^{-2\rho} {g^{(2)[1]}}_{zz} -\frac{1}{2} e^{-2\rho} {g^{(2)[1]}}_{\bar{z}\bar{z}} + {g^{(2)[1]}}_{z\bar{z}} \nonumber\\
    &+ \pi^2 (e^{2\rho} - e^{-2\rho}) g^{(0)[1]}_{z\bar{z}}, \nonumber\\
   {g^{FG[1]}}_{\bar{z}\bar{z}} =& {g^{(2)[1]}}_{\bar{z}\bar{z}} - e^{-2\rho} {g^{(2)[1]}}_{z\bar{z}} + \pi^2 e^{2\rho} g^{(0)[1]}_{\bar{z}\bar{z}} \nonumber\\
   &- \pi^2 e^{-2\rho} g^{(0)[1]}_{zz}.
\end{align}

We require the metric to be regular at $\rho = 0$, that is, its components in the $(t,x,y)$ coordinates
\begin{align}
    t &= t, \notag\\
    x &= \rho \cos\phi, \notag\\
    y &= \rho \sin\phi
\end{align}
which properly covers $\rho = 0$, are regular. We have the transformation equation of the components
\begin{align}
    g_{tt} &= g_{tt}, \nonumber\\
    g_{t\rho} &= g_{tx} \cos\phi + g_{ty} \sin\phi, \nonumber\\
    g_{t\phi} &=\rho(-g_{tx} \sin\phi + g_{ty} \cos\phi), \nonumber\\
    g_{\rho\rho} &= g_{xx}\cos^2\phi + 2g_{xy}\cos\phi\sin\phi + g_{yy} \sin^2\phi, \nonumber\\
    g_{\phi\phi} &= \rho^2(g_{xx}\sin^2\phi - 2g_{xy}\cos\phi\sin\phi + g_{yy} \cos^2\phi), \nonumber\\
    g_{\rho\phi} &= \rho[-(g_{xx}-g_{yy})\cos\phi\sin\phi + g_{xy}(\cos^2\phi-\sin^2\phi)].
\end{align}
Take the $\rho\to 0$ limit, the components in $(t,x,y)$ coordinates on the right-hand side should go to a limit that can only depend on $t$ with a period $\text{Im}\tau$
\begin{align}
    \lim_{\rho\to 0} g_{tt} =& g^*_{tt}(t), \nonumber\\
    \lim_{\rho\to 0} g_{t\rho} =& g^*_{tx}(t) \cos\phi + g^*_{ty}(t) \sin\phi, \nonumber\\
    \lim_{\rho\to 0} \frac{g_{t\phi}}{\rho} =&-g^*_{tx}(t) \sin\phi + g^*_{ty}(t) \cos\phi, \nonumber\\
    \lim_{\rho\to 0} g_{\rho\rho} =& g^*_{xx}(t)\cos^2\phi + 2g^*_{xy}(t)\cos\phi\sin\phi + g^*_{yy}(t) \sin^2\phi, \nonumber\\
    \lim_{\rho\to 0} \frac{g_{\phi\phi}}{\rho^2} =& g^*_{xx}(t)\sin^2\phi - 2g^*_{xy}(t)\cos\phi\sin\phi + g^*_{yy}(t) \cos^2\phi, \nonumber\\
    \lim_{\rho\to 0} \frac{g_{\rho\phi}}{\rho} =& -(g^*_{xx}(t)-g^*_{yy}(t))\cos\phi\sin\phi \nonumber\\
    &+ g^*_{xy}(t)(\cos^2\phi-\sin^2\phi).
\end{align}
Now we impose these conditions on the varied bulk metric in (\ref{Perturbed metric first order in epsilon}), and we find
\begin{widetext}
\begin{align} \label{Global regularity condition first order in epsilon}
    &\lim_{\rho\to 0} (2\cosh\rho\sinh\rho {V^{[1]}}^\rho + 2 \cosh^2\rho \partial_t {V^{[1]}}^t + g^{FG[1]}_{tt}) = {g^{*[1]}}_{tt}(t), \nonumber\\
    &\lim_{\rho\to 0} (\partial_t {V^{[1]}}^\rho + \cosh^2\rho\partial_\rho {V^{[1]}}^t) = {g^{*[1]}}_{tx}(t) \cos\phi + {g^{*[1]}}_{ty}(t) \sin\phi, \nonumber\\
    &\lim_{\rho\to 0} \frac{\cosh^2\rho \partial_\phi {V^{[1]}}^t + \sinh^2\rho \partial_t {V^{[1]}}^\phi + g^{FG[1]}_{t\phi}}{\rho} = -{g^{*[1]}}_{tx}(t)\sin\phi + {g^{*[1]}}_{ty}(t)\cos\phi, \nonumber\\
    &\lim_{\rho\to 0} 2\partial_\rho {V^{[1]}}^\rho = {g^{*[1]}}_{xx}(t)\cos^2\phi + 2{g^{*[1]}}_{xy}(t)\cos\phi\sin\phi + {g^{*[1]}}_{yy}(t)\sin^2\phi, \nonumber\\
    &\lim_{\rho\to 0} \frac{2\cosh\rho\sinh\rho {V^{[1]}}^\rho + 2\sinh^2\rho \partial_\phi {V^{[1]}}^\phi + g^{FG[1]}_{\phi\phi}}{\rho^2} = {g^{*[1]}}_{xx}(t)\sin^2\phi - 2{g^{*[1]}}_{xy}(t)\cos\phi\sin\phi + {g^{*[1]}}_{yy}(t)\cos^2\phi, \nonumber\\
    &\lim_{\rho\to 0} \frac{\partial_\phi {V^{[1]}}^\rho + \sinh^2\rho \partial_\rho {V^{[1]}}^\phi}{\rho} = -({g^{*[1]}}_{xx}(t)-{g^{*[1]}}_{yy}(t))\cos\phi\sin\phi + {g^{*[1]}}_{xy}(t)(\cos^2\phi-\sin^2\phi).
\end{align}
In addition, we have the power series expansion of (\ref{delta gFG1}) at $\rho= 0$
\begin{align} \label{Power series of gFG1 in rho}
    g^{FG[1]}_{\phi\phi} &= g^{FG[1]}_{1\phi\phi} \rho + g^{FG[1]}_{2\phi\phi} \rho^2 + \text{O}(\rho^3) = [\frac{1}{2\pi^2}({g^{(2)[1]}}_{zz}+2{g^{(2)[1]}}_{z\bar{z}}+{g^{(2)[1]}}_{\bar{z}\bar{z}})+g^{(0)[1]}_{zz}+2g^{(0)[1]}_{z\bar{z}}+g^{(0)[1]}_{\bar{z}\bar{z}}]\rho \nonumber\\
    &- \frac{1}{2\pi^2}({g^{(2)[1]}}_{zz}+2{g^{(2)[1]}}_{z\bar{z}}+A_{\bar{z}\bar{z}})\rho^2 + \text{O}(\rho^3), \nonumber\\
    g^{FG[1]}_{t\phi} &= g^{FG[1]}_{0t\phi} + \text{O}(\rho^2) = \frac{i}{4\pi^2}({g^{(2)[1]}}_{zz}-{g^{(2)[1]}}_{\bar{z}\bar{z}}) + \frac{i}{2}(g^{(0)[1]}_{zz}-g^{(0)[1]}_{\bar{z}\bar{z}}) + \text{O}(\rho^2), \nonumber\\
    g^{FG[1]}_{tt} &= g^{FG[1]}_{0tt} + \text{O}(\rho) = -\frac{1}{2\pi^2}({g^{(2)[1]}}_{zz}-2{g^{(2)[1]}}_{z\bar{z}}+{g^{(2)[1]}}_{\bar{z}\bar{z}}) + \text{O}(\rho).
\end{align}
\end{widetext}
Integrating the third equation in (\ref{Global regularity condition first order in epsilon}) over the torus, we find
\begin{align} \label{Global regularity condition for C1 no1}
    \int_{\text{T}^2} d^2z \: g^{FG[1]}_{0t\phi} = 0.
\end{align}
From the fourth equation in (\ref{Global regularity condition first order in epsilon}) we know ${V^{[1]}}^\rho$ can be linearly approximated near $\rho=0$
\begin{align}
    {V^{[1]}}^\rho = a_0 + a_1 \rho + \text{o}(\rho).
\end{align}
Then we subtract the fifth equation from the fourth and integrate it over the torus. We find
\begin{align} \label{Global regularity condition for C1 no2}
    \int_{\text{T}^2} d^2z \: g^{FG[1]}_{2\phi\phi} = 0.
\end{align}
Plugging (\ref{Power series of gFG1 in rho}) into (\ref{Global regularity condition for C1 no1}) and (\ref{Global regularity condition for C1 no2}), we obtain the global regularity condition (\ref{global regularity condition thermal AdS3}).

\section{Computation of holographic correlators for $T\bar{T}$-deformed CFT} \label{C}
As in the section \ref{sec:conformal boundary}, to compute holographic correlators we solve the varied one-point correlator order by order from (\ref{one-point correlator conservation}) and (\ref{one-point correlator trace relation}). For the first order, we get
\begin{widetext}
\begin{align}
    &\partial_{\bar{z}} {T^{[1]}}_{zz} + \partial_z {T^{[1]}}_{z\bar{z}} = \frac{\pi}{8G} \frac{1-e^{-2\rho_c}}{1+e^{-2\rho_c}} (-\partial_z f_{zz} - 3 \partial_z f_{\bar{z}\bar{z}} + 2e^{-2\rho_c}\partial_{\bar{z}}f_{zz} + 2e^{-2\rho_c}\partial_z f_{z\bar{z}}), \label{cutoff two point eqn1}\\
    &\partial_{z} {T^{[1]}}_{\bar{z}\bar{z}} + \partial_{\bar{z}} {T^{[1]}}_{z\bar{z}} = \frac{\pi}{8G} \frac{1-e^{-2\rho_c}}{1+e^{-2\rho_c}} (-\partial_{\bar{z}} f_{\bar{z}\bar{z}} - 3 \partial_{\bar{z}} f_{zz} + 2e^{-2\rho_c}\partial_z f_{\bar{z}\bar{z}} + 2e^{-2\rho_c}\partial_{\bar{z}} f_{z\bar{z}}), \label{cutoff two point eqn2}
\end{align}
and
\begin{align}
    &(1+e^{-4\rho_c}){T^{[1]}}_{z\bar{z}} + e^{-2\rho_c}({T^{[1]}}_{zz}+{T^{[1]}}_{\bar{z}\bar{z}}) \nonumber\\
    &= -\frac{\pi(1-e^{-2\rho_c})}{8G}^2(f_{zz}+f_{\bar{z}\bar{z}}+2e^{-2\rho_c}f_{z\bar{z}}) + \frac{1-e^{-4\rho_c}}{16\pi G}(\partial_{\bar{z}}^2 f_{zz} - 2\partial_z\partial_{\bar{z}}f_{z\bar{z}} + \partial_z^2 f_{\bar{z}\bar{z}}). \label{cutoff two point eqn3}
\end{align}
Adding $\partial_z$(\ref{cutoff two point eqn1}) and $\partial_{\bar{z}}$(\ref{cutoff two point eqn2}), and then plugging in (\ref{cutoff two point eqn3}), we get an equation of ${T^{[1]}}_{z\bar{z}}$
\begin{align}
    &(1-e^{-2\rho_c})^2 \partial_Z \partial_{\bar{Z}} {T_1}_{z\bar{z}} = -\frac{\pi(1-e^{-2\rho_c})^2}{8G} \partial_z\partial_{\bar{z}}(f_{zz} + f_{\bar{z}\bar{z}} + 2e^{-2\rho_c}f_{z\bar{z}}) + \frac{1-e^{-4\rho_c}}{16\pi G}\partial_z\partial_{\bar{z}} (\partial_{\bar{z}}^2 f_{zz} - 2\partial_z\partial_{\bar{z}}f_{z\bar{z}} + \partial_z^2 f_{\bar{z}\bar{z}}) \notag\\
    &-\frac{\pi e^{-2\rho_c}}{8G} \frac{1-e^{-2\rho_c}}{1+e^{-2\rho_c}}\Big[-\partial_z^2 f_{zz} - 3\partial_z^2 f_{\bar{z}\bar{z}} - \partial_{\bar{z}}^2 f_{\bar{z}\bar{z}} - 3\partial_{\bar{z}}^2 f_{zz} + 2e^{-2\rho_c}(\partial_z\partial_{\bar{z}} f_{zz} + \partial_z^2 f_{z\bar{z}} + \partial_z\partial_{\bar{z}}f_{\bar{z}\bar{z}} + \partial_{\bar{z}}^2 f_{z\bar{z}})\Big].
\end{align}
We solve this equation with the Green's function $\tilde{G}_\Omega$ on torus (see the appendix \ref{A})
\begin{align}
    &{T^{[1]}}_{z\bar{z}}(z) = \frac{1}{8\pi^2 G} \int_{\text{T}^2} d^2W \tilde{G}_\Omega(Z-W) \Big[-\pi^2 \partial_w\partial_{\bar{w}}(f_{zz} + f_{\bar{z}\bar{z}} + 2e^{-2\rho_c}f_{z\bar{z}}) + \frac{1+e^{-2\rho_c}}{2(1-e^{-2\rho_c})}\partial_w\partial_{\bar{w}} (\partial_{\bar{w}}^2 f_{zz} - 2\partial_w\partial_{\bar{w}}f_{z\bar{z}} + \partial_w^2 f_{\bar{z}\bar{z}}) \notag\\
    &-\frac{\pi^2 e^{-2\rho_c}}{1-e^{-4\rho_c}} (-\partial_w^2 f_{zz} - 3\partial_w^2 f_{\bar{z}\bar{z}} - \partial_{\bar{w}}^2 f_{\bar{z}\bar{z}} - 3\partial_{\bar{w}}^2 f_{zz} + 2e^{-2\rho_c}(\partial_w\partial_{\bar{w}} f_{zz} + \partial_w^2 f_{z\bar{z}} + \partial_w\partial_{\bar{w}}f_{\bar{z}\bar{z}} + \partial_{\bar{w}}^2 f_{z\bar{z}}))
    \Big] (w) + \frac{1}{8\pi G} D^{[1]},
\end{align}
and ${T^{[1]}}_{zz},{T^{[1]}}_{\bar{z}\bar{z}}$ follow from (\ref{cutoff two point eqn2}) and (\ref{cutoff two point eqn3})
\begin{align}
    {T^{[1]}}_{zz} = &\frac{1}{\pi} \int_{\text{T}^2} d^2w G_\tau(z-w) \Big[\frac{\pi}{8G}\frac{1-e^{-2\rho_c}}{1+e^{-2\rho_c}}(-\partial_w f_{zz} - 3 \partial_w f_{\bar{z}\bar{z}} + 2e^{-2\rho_c}\partial_{\bar{w}}f_{zz} + 2e^{-2\rho_c}\partial_w f_{z\bar{z}}) - \partial_w {T^{[1]}}_{z\bar{z}} \Big] + \frac{E^{[1]}}{8\pi G}, \\
    {T^{[1]}}_{\bar{z}\bar{z}} = &\frac{1}{\pi} \int_{\text{T}^2} d^2w \overline{G_\tau(z-w)} \Big[\frac{\pi}{8G}\frac{1-e^{-2\rho_c}}{1+e^{-2\rho_c}}(-\partial_{\bar{w}} f_{\bar{z}\bar{z}} - 3 \partial_{\bar{w}} f_{zz} + 2e^{-2\rho_c}\partial_w f_{\bar{z}\bar{z}} + 2e^{-2\rho_c}\partial_{\bar{w}} f_{z\bar{z}}) - \partial_{\bar{w}} {T^{[1]}}_{z\bar{z}} \Big] + \frac{\bar{E}^{[1]}}{8\pi G},
\end{align}
where $D^{[1]},E^{[1]},\bar{E}^{[1]}$ are constants of integration. The global regularity condition reads in the present context as
\begin{align}
    &\int_{\text{T}^2} d^2Z \Big[ {g^{(2)[1]}}_{ZZ}-{g^{(2)[1]}}_{\bar{Z}\bar{Z}} + 2\pi^2 (1-e^{-2\rho_c})^2 ({g^{(0)[1]}}_{ZZ} - {g^{(0)[1]}}_{\bar{Z}\bar{Z}}) \Big] = 0, \label{cutoff regularity condition 1}\\
    &\int_{\text{T}^2} d^2Z \Big[{g^{(2)[1]}}_{ZZ} + 2{g^{(2)[1]}}_{Z\bar{Z}} + {g^{(2)[1]}}_{\bar{Z}\bar{Z}}\Big] =0. \label{cutoff regularity condition 2}
\end{align}
In addition, by integrating (\ref{cutoff two point eqn3}) over the torus we find
\begin{align}
    &(1+e^{-4\rho_c})D^{[1]} + e^{-2\rho_c}(E^{[1]}+\bar{E}^{[1]}) = -\pi^2(1-e^{-2\rho_c})^2 \frac{1}{\text{Im}\tau} \int_{\text{T}^2}d^2z (f_{zz}+f_{\bar{z}\bar{z}}+2e^{-2\rho_c}f_{z\bar{z}}).
\end{align}
We determine the constants from the three equations above
\begin{align}
    D^{[1]} =& -\frac{\pi^2(1-e^{-2\rho_c})}{(1+e^{-2\rho_c})^3}\Big[(1+4e^{-2\rho_c}+e^{-4\rho_c})\frac{1}{\text{Im}\tau}\int_{\text{T}} d^2z (f_{zz}+f_{\bar{z}\bar{z}}) + 2e^{-2\rho_c}(1-2e^{-2\rho_c}-e^{-4\rho_c})\frac{1}{\text{Im}\tau}\int_{\text{T}} d^2z f_{z\bar{z}}\Big], \notag\\
    E^{[1]} =&\frac{2\pi^2(1-e^{-2\rho_c})}{(1+e^{-2\rho_c})^3}\Big[(1+e^{-2\rho_c}+e^{-4\rho_c})\frac{1}{\text{Im}\tau}\int_{\text{T}^2} d^2z f_{\bar{z}\bar{z}} + (2e^{-4\rho_c}+e^{-6\rho_c}) \frac{1}{\text{Im}\tau}\int_{\text{T}^2} d^2z f_{zz} - 2e^{-2\rho_c}\frac{1}{\text{Im}\tau}\int_{\text{T}^2} d^2z f_{z\bar{z}}\Big], \notag\\
    \bar{E}^{[1]} =&\frac{2\pi^2(1-e^{-2\rho_c})}{(1+e^{-2\rho_c})^3}\Big[(1+e^{-2\rho_c}+e^{-4\rho_c})\frac{1}{\text{Im}\tau}\int_{\text{T}^2} d^2z f_{zz} + (2e^{-4\rho_c}+e^{-6\rho_c}) \frac{1}{\text{Im}\tau}\int_{\text{T}^2} d^2z f_{\bar{z}\bar{z}} - 2e^{-2\rho_c}\frac{1}{\text{Im}\tau}\int_{\text{T}^2} d^2z f_{z\bar{z}}\Big].
\end{align}
We can compute all two-point correlators by taking the functional derivative of $T^{[1]}_{ij}$ with respect to $f_{ij}$. As in section \ref{sec:conformal boundary}, the constants can also be determined from (\ref{global metric variation and modular differentiation eqn 1}) and (\ref{global metric variation and modular differentiation eqn 2}). Setting $O=T_{zz}$ we obtain
\begin{align}
    (\bar{\tau}-\tau) \partial_\tau \langle T_{zz}(w) \rangle &= (w-\bar{w})\partial_w \langle T_{zz}(w) \rangle + 2 \langle T_{zz}(w) \rangle 
    + \int_{\text{T}^2} d^2z \big( \frac{\delta \langle T_{zz}(w) \rangle}{\delta \gamma_{\bar{z}\bar{z}}(z)} -  \frac{\delta \langle T_{zz}(w) \rangle}{\delta \gamma_{z\bar{z}}(z)} \big),\nonumber\\
     (\tau-\bar{\tau}) \partial_{\bar{\tau}} \langle T_{zz}(w) \rangle &= (\bar{w}-w)\partial_{\bar{w}} \langle T_{zz}(w) \rangle - 2 \langle T_{z\bar{z}}(w) \rangle 
     + \int_{\text{T}^2} d^2z \big( \frac{\delta \langle T_{zz}(w) \rangle}{\delta \gamma_{zz}(z)} -  \frac{\delta \langle T_{zz}(w) \rangle}{\delta \gamma_{z\bar{z}}(z)} \big).
\end{align}
In addition, we take a functional derivative with respect to $\gamma_{\bar{z}\bar{z}}$ in (\ref{one-point correlator trace relation}), and integrate over one point
\begin{align}
    \int_{\text{T}^2} d^2z \frac{\delta \langle T_{z\bar{z}}(z)\rangle}{\delta \gamma_{\bar{z}\bar{z}}(w)} - \langle T_{zz}(w) \rangle + 8\pi G r_c^2 \int_{\text{T}^2} d^2z \Big[2\langle T_{z\bar{z}}(z)\rangle \frac{\delta \langle T_{z\bar{z}}(z)\rangle}{\delta \gamma_{\bar{z}\bar{z}}(w)} - \langle T_{zz}(z)\rangle \frac{\delta \langle T_{\bar{z}\bar{z}}(z)\rangle}{\delta \gamma_{\bar{z}\bar{z}}(w)} - \langle T_{\bar{z}\bar{z}}(z)\rangle \frac{\delta \langle T_{zz}(z)\rangle}{\delta \gamma_{\bar{z}\bar{z}}(w)} \Big] = 0.
\end{align}
The three equations above determine the constants $D^{[1]},E^{[1]},\bar{E}^{[1]}$ as one-point-averaged correlators.

A recurrence algorithm for holographic correlators similar to (\ref{recurrence relation conformal boundary}) can be derived, but it takes a much more complicated form. Turning on a variation of the metric $\gamma_{\bar{z}\bar{z}}=F$, the varied one-point correlator is solved as 
\begin{align}\label{T_zzbar^[n]}
    T_{z\bar{z}}^{[n]}(z) = 
    & \frac{1}{\pi}\int_{\text{T}^2} d^2 W \tilde{G}_{\Omega}(Z-W) \Big(  -\frac{e^{-2 \rho_c}}{ ( 1 - e^{-2 \rho_c} )^2 } \big( \partial ( 3 \partial F T_{zz}^{[n-1]} + 2 F \partial T_{zz}^{[n-1]} ) +\bar{\partial} ( \bar{\partial} F T_{zz}^{[n-1]} ) +2\partial\bar{\partial} (F T_{z\bar{z}}^{[n-1]} ) \big) \notag\\
    &\qquad +\frac{ 1 - e^{-4\rho_c} }{( 1 - e^{-2 \rho_c} )^2} \partial\bar{\partial} \big( F T_{zz}^{[n-1]} +\frac{\partial^2 F}{16 \pi G} - 8 \pi G r_c^2 \sum_{m=1}^{n-1} ( T_{z\bar{z}}^{[m]} T_{z\bar{z}}^{[n-m]} - T_{zz}^{[m]} T_{\bar{z}\bar{z}}^{[n-m]} ) \big) \Big)(w) +\frac{D^{[n]}}{8\pi G},
\end{align}
and 
\begin{align}
    T_{zz}^{[n]}(z) =& \frac{1}{\pi} \int_{\text{T}^2} d^2 w G_{\tau}(z-w) ( -\partial T_{z\bar{z}}^{[n]} + 3 \partial F T_{zz}^{[n-1]} + 2 F \partial T_{zz}^{[n-1]} )(w) +\frac{E^{[n]}}{8\pi G}, \label{T_zz^[n]}\\
    T_{\bar{z}\bar{z}}^{[n]}(z) =& \frac{1}{\pi} \int_{\text{T}^2} d^2 w \overline{G_{\tau}(z-w)} \big( -\bar{\partial} T_{z\bar{z}}^{[n]} +\bar{\partial} F T_{zz}^{[n-1]} +2\partial(F T_{z\bar{z}}^{[n-1]}) \big)(w) +\frac{\bar{E}^{[n]}}{8\pi G}, \label{T_zbarzbar^[n]}
\end{align}
where $D^{[n]}$, $E^{[n]}$ and $\bar{E}^{[n]}$ are constants of integration. To fix $D^{[n]}$, $E^{[n]}$ and $\bar{E}^{[n]}$, we set $O=T_{zz}(z_1)\ldots T_{zz}(z_n)$ in (\ref{global metric variation and modular differentiation eqn 1}) and (\ref{global metric variation and modular differentiation eqn 2}), and obtain
\begin{align}
    (\bar{\tau}-\tau) \partial_\tau \langle T_{zz}(z_1)\ldots T_{zz}(z_n) \rangle = 
    &\sum_{i=1}^{n}[(z_i -\bar{z_i}) \partial_{z_i} \langle T_{zz}(z_1)\ldots T_{zz}(z_n) \rangle +2 \langle T_{zz}(z_1)\ldots T_{zz}(z_n) \rangle] \notag\\
    & +\int_{\text{T}^2} d^2z [\frac{\delta \langle T_{zz}(z_1)\ldots T_{zz}(z_n) \rangle}{\delta \gamma_{\bar{z}\bar{z}}(z)} - \frac{\delta \langle T_{zz}(z_1)\ldots T_{zz}(z_n) \rangle}{\delta \gamma_{z\bar{z}}(z)} ], \label{TTbar global metric variation and modular differentiation eqn 1} \\
    (\tau-\bar{\tau}) \partial_{\bar{\tau}} \langle T_{zz}(z_1)\ldots T_{zz}(z_n) \rangle = 
    & \sum_{i=1}^{n}[(\bar{z}_i -z_i)\partial_{\bar{z}_i} \langle T_{zz}(z_1)\ldots T_{zz}(z_n) \rangle -2 \langle T_{z\bar{z}}(z_i) T_{zz}(z_1)\ldots\hat{T}_{zz}(z_i)\ldots T_{zz}(z_n)] \notag\\
    & +\int_{\text{T}^2} d^2z [\frac{\delta \langle T_{zz}(z_1)\ldots T_{zz}(z_n) \rangle}{\delta \gamma_{zz}(z)} - \frac{\delta \langle T_{zz}(z_1)\ldots T_{zz}(z_n) \rangle}{\delta \gamma_{z\bar{z}}(z)} ], \label{TTbar global metric variation and modular differentiation eqn 2}
\end{align}
where $\hat{T}_{zz}(z_i)$ means dropping the $i-$th operator.
In addition, we take n times of functional derivatives with respect to $\gamma_{\bar{z}\bar{z}}(z_1),\ldots,\gamma_{\bar{z}\bar{z}}(z_n)$ in (\ref{one-point correlator trace relation}), and integrate over one point to get
\begin{align}
    \int_{\text{T}^2} d^2z \langle T_{z\bar{z}}(z) T_{zz}(z_1)\ldots T_{zz}(z_n) \rangle =
    & -\frac{e^{-2\rho_c}}{1+ e^{-4\rho_c}}\big(\int_{\text{T}^2} d^2z \langle T_{zz}(z) T_{zz}(z_1)\ldots T_{zz}(z_n) \rangle +\int_{\text{T}^2} d^2z \langle T_{\bar{z}\bar{z}}(z) T_{zz}(z_1)\ldots T_{zz}(z_n) \rangle\big) \notag\\
    & +\frac{1-e^{-4\rho_c}}{1+e^{-4\rho_c}}\big(\frac{n}{2}\langle T_{zz}(z_1)\ldots T_{zz}(z_n) \rangle -\int_{\text{T}^2} d^2z M(z,z_1,\ldots,z_n)\big), \label{TTbar integration relation}
\end{align}
where we denote
\begin{align}
    M(z,z_1,\ldots,z_n) 
    = & \sum_{m=1}^{n-1} \sum_{\sigma}\frac{8 \pi G r_c^2}{m!(n-m)!} \Big[ \langle T_{z\bar{z}}(z) T_{zz}(z_{\sigma(1)}) \ldots T_{zz}(z_{\sigma(m)}) \rangle \langle T_{z\bar{z}}(z) T_{zz}(z_{\sigma(m+1)}) \ldots T_{zz}(z_{\sigma(n)}) \rangle \notag\\
    & - \langle T_{zz}(z) T_{zz}(z_{\sigma(1)}) \ldots T_{zz}(z_{\sigma(m)}) \rangle \langle T_{\bar{z}\bar{z}}(z) T_{zz}(z_{\sigma(m+1)}) \ldots T_{zz}(z_{\sigma(n)}) \rangle \Big]
\end{align}
with $\sigma$ running over all permutations of $(1,\ldots,n)$.

With the constants of integration determined by (\ref{TTbar global metric variation and modular differentiation eqn 1}), (\ref{TTbar global metric variation and modular differentiation eqn 2}) and (\ref{TTbar integration relation}), we obtain the following recurrence relations for correlators
\begin{align}
    & \langle T_{z\bar{z}}(z) T_{zz}(z_{1}) \ldots T_{zz}(z_{n}) \rangle \notag\\
    = & -\frac{1}{\pi}\frac{1}{1-e^{-4\rho_c}} \sum_{i=1}^{n} \Big\{ \Big[ \frac{e^{-2\rho_c}}{2(1+e^{-2\rho_c})} \big( (\partial_{Z_i} +e^{-2\rho_c}\partial_{\bar{Z}_i})\tilde{G}_\Omega(Z-Z_i)\partial_{z_i} + (e^{-2\rho_c}\partial_{Z_i}+\partial_{\bar{Z}_i}) \tilde{G}_\Omega(Z-Z_i) \partial_{\bar{z}_i} \big) \notag\\ 
    & -\frac{1}{2(1+e^{-2\rho_c})^2} \big( -2(e^{-2\rho_c}+e^{-6\rho_c})\partial_Z^2 -4e^{-6\rho_c}\partial_{\bar{Z}}^2\big) \tilde{G}_\Omega(Z-Z_i)\Big] \langle T_{zz}(z_{1}) \ldots T_{zz}(z_{n}) \rangle \notag\\
    & +\frac{e^{-2\rho_c}}{(1+e^{-2\rho_c})^2} (e^{-2\rho_c}\partial_Z^2 +e^{-2\rho_c}\partial_{\bar{Z}}^2) \tilde{G}_\Omega(Z-Z_i) \langle T_{z\bar{z}}(z_i) T_{zz}(z_{1}) \ldots  \hat{T}_{zz}(z_i) \ldots T_{zz}(z_{n}) \rangle \Big\}\notag\\
    &+\sum_{i=1}^n\big[\frac{1-8e^{-4\rho_c}-e^{-8\rho_c}}{2(1-e^{-4\rho_c})^2}\langle T_{zz}(z_{1})  \ldots T_{zz}(z_{n}) \rangle-\frac{e^{-2\rho_c}+e^{-6\rho_c}}{(1-e^{-4\rho_c})^2}\langle T_{z\bar{z}}(z_i) T_{zz}(z_{1}) \ldots  \hat{T}_{zz}(z_i) \ldots T_{zz}(z_{n}) \rangle\big]\big(\delta(z-z_i)-\frac{1}{\text{Im}\tau}\big)\notag\\
    & +\frac{\delta_{n,1}}{32\pi^2 G}\Big\{ \frac{e^{-2\rho_c}}{(1+e^{-2\rho_c})^4}(\partial_Z^4 +e^{-4\rho_c}\partial_{\bar{Z}}^4)\tilde{G}_\Omega(Z-Z_1) +\frac{\pi}{(1-e^{-4\rho_c})^3}\Big[ (1+e^{-12\rho_c})\partial_z^2 +(e^{-4\rho_c} +e^{-8\rho_c})\partial_{\bar{z}}^2 \notag\\
    & +(e^{-2\rho_c} -6e^{-6\rho_c} +e^{-10\rho_c})\partial_z \partial_{\bar{z}} \Big]\delta(z-z_1) \Big\} -\frac{1}{\pi}\int_{\text{T}^2} d^2w\tilde G_{\Omega}(Z-W) \partial_{w} \partial_{\bar w} M(w,z_1,\ldots,z_n) \notag\\
    & + \frac{1}{(1 + e^{-2\rho_c})^2 \text{Im}\tau } \Big\{-(1-e^{-4\rho_c})\int_{\text{T}^2} d^2 w M(w,z_1,\ldots,z_n)-e^{-2\rho_c}\sum_{i=1}^{n} \langle T_{z\bar{z}}(z_i) T_{zz}(z_{1}) \ldots \hat{T}_{zz}(z_i) \ldots T_{zz}(z_{n}) \rangle  \notag\\
    & - \frac{1}{2}e^{-2\rho_c} \Big[(\bar{\tau}-\tau) \partial_\tau +(\tau-\bar{\tau}) \partial_{\bar{\tau}}+\sum_{i=1}^{n}(\bar{z}_i-z_i) \partial_{z_i}+ \sum_{i=1}^{n}(z_i-\bar{z_i})\partial_{\bar{z}_i} -n(4+e^{2\rho_c}-e^{-2\rho_c}) \Big] \langle T_{zz}(z_1) \ldots T_{zz}(z_n)\rangle \Big\}.
\end{align} 
\begin{align}
    &\langle T_{zz}(z) T_{zz}(z_{1}) \ldots T_{zz}(z_{n}) \rangle \notag\\
    =& \frac{1}{\pi} \frac{1}{1-e^{-4\rho_c}} \sum_{i=1}^{n} \Big\{ \Big[ -\frac{1}{2(1+e^{-2\rho_c})}\big((\partial_Z +e^{-6\rho_c}\partial_{\bar{Z}}) \tilde{G}_\Omega(Z-Z_i) \partial_{z_i} +(e^{-2\rho_c}\partial_Z +e^{-4\rho_c}\partial_{\bar{Z}}) \tilde{G}_\Omega(Z-Z_i)\partial_{\bar{z}_i}\big) \notag\\
    & +\frac{1}{(1+e^{-2\rho_c})^2}\big((1+e^{-4\rho_c})\partial_Z^2 +2 e^{-8\rho_c}\partial_{\bar{Z}}^2\big) \tilde{G}_\Omega(Z-Z_i) \Big] \langle T_{zz}(z_{1}) \ldots T_{zz}(z_{n}) \rangle \notag\\
    & +\frac{1}{(1+e^{-2\rho_c})^2} (e^{-2\rho_c} \partial_Z^2 +e^{-6\rho_c}\partial_{\bar{Z}}^2) \tilde{G}_\Omega(Z-Z_i) \langle T_{z\bar{z}}(z_i) T_{zz}(z_{1}) \ldots  \hat{T}_{zz}(z_i) \ldots T_{zz}(z_{n}) \rangle \Big\} \notag\\
    & -\frac{\delta_{n,1}}{32\pi^2 G} \frac{1}{(1+e^{-2\rho_c})^4} (\partial_Z^4 +e^{-8\rho_c}\partial_{\bar{Z}}^4) \tilde{G}_\Omega(Z-Z_1) + \frac{1}{\pi} \int_{\text{T}^2} d^2 w \tilde{G}_\Omega(Z-W) \partial_{w}^2M(w,z_1,\ldots,z_n) \notag\\
    & +\sum_{i=1}^{n} \Big[\frac{e^{-2\rho_c} +3e^{-6\rho_c}}{(1 -e^{-4\rho_c})^2} \langle T_{zz}(z_{1}) \ldots T_{zz}(z_{n}) \rangle +\frac{2 e^{-4\rho_c}}{(1-e^{-4\rho_c})^2} \langle T_{z\bar{z}}(z_i) T_{zz}(z_{1}) \ldots  \hat{T}_{zz}(z_i) \ldots T_{zz}(z_{n}) \rangle \Big] \big(\delta(z-z_i) -\frac{1}{\text{Im}\tau}\big)  \notag\\
    & -\frac{\delta_{n,1}}{16\pi G}\frac{1}{(1-e^{-4\rho_c})^3}\Big[(2e^{-2\rho_c} -3e^{-6\rho_c} +2e^{-10\rho_c})\partial_z^2 -(e^{-4\rho_c}+e^{-8\rho_c})\partial_z \partial_{\bar{z}} +e^{-6\rho_c}\partial_{\bar{z}}^2\Big] \delta(z-z_1) \notag\\
    &+ \frac{1}{(1 + e^{-2\rho_c})^2 \text{Im}\tau } \Big\{  -(1-e^{-4\rho_c})\int_{\text{T}^2} d^2 w M(w,z_1,\ldots ,z_n)-e^{-2\rho_c} \sum_{i=1}^{n} \langle T_{z\bar{z}}(z_i) T_{zz}(z_{1}) \ldots  \hat{T}_{zz}(z_i) \ldots T_{zz}(z_{n}) \rangle \notag\\
    & +\frac{1}{2}\Big[(1+e^{-2\rho_c}+e^{-4\rho_c}) (\bar{\tau}-\tau) \partial_\tau -e^{-2\rho_c}(\tau-\bar{\tau}) \partial_{\bar{\tau}} +(1+e^{-2\rho_c}+e^{-4\rho_c}) \sum_{i=1}^{n}(\bar{z}_i-z_i) \partial_{z_i} -e^{-2\rho_c}\sum_{i=1}^{n}(z_i-\bar{z_i})\partial_{\bar{z}_i} \notag\\
    &-2n(1+e^{-2\rho_c}+2e^{-4\rho_c}) \Big] \langle T_{zz}(z_1) \ldots T_{zz}(z_n)\rangle \Big\} .
\end{align}
\begin{align}
    &\langle T_{\bar{z}\bar{z}}(z) T_{zz}(z_{1}) \ldots T_{zz}(z_{n}) \rangle \notag\\
    =& \frac{1}{\pi}\frac{1}{1-e^{-4\rho_c}} \sum_{i=1}^{n} \Big\{ \Big[ -\frac{1}{2(1+e^{-2\rho_c})} \big( (e^{-4\rho_c}\partial_{Z} +e^{-2\rho_c}\partial_{\bar{Z}})\tilde{G}_{\Omega}(Z-Z_i)\partial_{z_i} +(e^{-6\rho_c}\partial_{Z}  +\partial_{\bar{Z}}) \tilde{G}_{\Omega}(Z-Z_i)\partial_{\bar{z}_i} \big)  \notag\\
    & + \frac{1}{(1+e^{-2\rho_c})^2}\big((e^{-4\rho_c}+e^{-8\rho_c})\partial_Z^2 +2 e^{-4\rho_c}\partial_{\bar{Z}}^2\big) \tilde{G}_{\Omega}(Z-Z_i) \Big] \langle T_{zz}(z_{1}) \ldots T_{zz}(z_{n}) \rangle \notag\\
    & +\frac{1}{(1+e^{-2\rho_c})^2}(e^{-6\rho_c}\partial_Z^2 +e^{-2\rho_c} \partial_{\bar{Z}}^2) \tilde{G}_{\Omega}(Z-Z_i) \langle T_{z\bar{z}}(z_i) T_{zz}(z_{1}) \ldots  \hat{T}_{zz}(z_i) \ldots T_{zz}(z_{n}) \rangle \Big\} \notag\\
    & -\frac{\delta_{n,1}}{32\pi^2 G} \frac{e^{-4\rho_c}}{(1+e^{-2\rho_c})^4}(\partial_Z^4 +\partial_{\bar{Z}}^4) \tilde{G}_{\Omega}(Z-Z_1)  +\frac{1}{\pi} \int_{\text{T}^2} d^2 w \tilde{G}_{\Omega}(Z-W) \partial_{\bar{w}}^2 M(w,z_1,\ldots,z_n)\notag\\
    &+\sum_{i=1}^{n}\Big[\frac{e^{-2\rho_c}+3e^{-6\rho_c}}{(1-e^{-4\rho_c})^2} \langle T_{zz}(z_{1}) \ldots T_{zz}(z_{n}) \rangle -\frac{1-4 e^{-4\rho_c}+e^{-8\rho_c}}{(1-e^{-4\rho_c})^2} \langle T_{z\bar{z}}(z_i) T_{zz}(z_{1}) \ldots  \hat{T}_{zz}(z_i) \ldots T_{zz}(z_{n}) \rangle \Big]\big(\delta(z-z_i)-\frac{1}{\text{Im}\tau}\big) \notag\\
    & -\frac{\delta_{n,1}}{32\pi G} \frac{1}{(1-e^{-4\rho_c})^3} \Big[ (e^{-2\rho_c}+e^{-10\rho_c})(\partial_z^2 +\partial_{\bar{z}}^2) +(1 -3e^{-4\rho_c} -3e^{-8\rho_c} +e^{-12\rho_c})\partial_z \partial_{\bar{z}} \Big]\delta(z-z_1) \notag\\
    &+ \frac{1}{(1 + e^{-2\rho_c})^2 \text{Im}\tau } \Big\{  -(1-e^{-4\rho_c})\int_{\text{T}^2} d^2 w M(w,z_1,...,z_n) + (1+e^{-2\rho_c}+e^{-4\rho_c}) \sum_{i=1}^{n} \langle T_{z\bar{z}}(z_i) T_{zz}(z_{1}) ...  \hat{T}_{zz}(z_i) ... T_{zz}(z_{n}) \rangle \notag\\
    & +\frac{1}{2} \Big[(1+e^{-2\rho_c}+e^{-4\rho_c})(\tau-\bar{\tau}) \partial_{\bar{\tau}} -e^{-2\rho_c} (\bar{\tau}-\tau) \partial_\tau + (1+e^{-2\rho_c}+e^{-4\rho_c}) \sum_{i=1}^{n} (z_i-\bar{z_i}) \partial_{\bar{z}_i} - e^{-2\rho_c} \sum_{i=1}^{n}(\bar{z}_i-z_i) \partial_{z_i} \notag\\ 
    & +2n(e^{-2\rho_c} -e^{-4\rho_c}) \Big] \langle T_{zz}(z_1) \ldots T_{zz}(z_n)\rangle \Big\}. 
\end{align}
By taking the limit $\rho_c\to\infty$, the second one reproduces the recurrence relation (\ref{recurrence relation conformal boundary}) for holographic CFTs.
\end{widetext}


\bibliography{reference.bib}

\end{document}